\documentclass[1p,review,number]{elsarticle}

\usepackage[latin1]{inputenc}
\usepackage[T1]{fontenc}

\usepackage{amsmath,amssymb}
\usepackage{graphicx}
\usepackage{dcolumn}
\usepackage{bm}
\graphicspath{{fig/}}

\usepackage[english]{babel}
\usepackage{units}
\usepackage{journames}

\usepackage[highlight]{neel}

\begin{document}

\begin{frontmatter}

\renewcommand\topfraction{0.8}
\renewcommand\bottomfraction{0.7}
\renewcommand\floatpagefraction{0.7}

\def\Htc{H_{t\mathrm{,c}}}

\title{Magnetostatics of synthetic ferrimagnet elements}%

\author[neel]{Olivier Fruchart\corref{cor1}}
\ead{Olivier.Fruchart@grenoble.cnrs.fr}

\author[spintec]{Bernard Di\'{e}ny}

\cortext[cor1]{Corresponding author}

\address[neel]{Institut N\'{E}EL, CNRS \& Universit\'{e} Joseph Fourier -- BP166 -- F-38042 Grenoble Cedex 9 -- France}%
\address[spintec]{SPINTEC (UMR8191 CEA/CNRS/UJF/G-INP), CEA Grenoble, INAC, 38054 Grenoble Cedex 9, France}%

\date{\today}%

\begin{abstract}

We calculate the magnetostatic energy of synthetic ferrimagnet~(SyF) elements, consisting of two thin
ferromagnetic layers coupled antiferromagnetically, \eg through RKKY coupling. Uniform magnetization is assumed in each layer. Exact formulas as well as approximate yet accurate ones are provided. These may be used to evaluate various quantities of SyF such as shape-induced coercivity and thermal stability, like demagnetizing coefficients are used in single elements.

\end{abstract}

\end{frontmatter}



\vskip 0.5in

\vskip 0.5in


\newcommand{\SyF}{SyF\/\xspace}
\newcommand{\SyFs}{SyFs\/\xspace}
\newcommand{\SAF}{SAF\/\xspace}
\newcommand{\SAFs}{SAFs\/\xspace}

Synthetic antiferromagnets~(SAF, resp. ferrimagnets, SyF)\cite{bib-HEI1993,bib-VAN1997} consist of two
thin ferromagnetic films of moments of same (resp. different) magnitude, strongly coupled
antiferromagnetically thanks to the RKKY interaction through an ultrathin spacer layer, typically
Ru $\thicknm{0.6-0.9}$ thick\cite{bib-PAR1991}. Hereon we  consider only the case of in-plane
magnetized layers. SyFs are widely used to provide spin-polarized layers displaying an overall weak
moment. One benefit is to minimize cross-talk of neighboring~(\eg memory bits) or stacked~(\eg in a
spin-valve) elements through stray-field coupling\cite{bib-HEI1993,bib-VAN1997}, such as in Magnetic
Random Access Memory~(MRAM)\cite{bib-SOU2005}. SyFs are also used to decrease the Zeeman coupling with
external fields, \eg to increase coercivity in reference layers\cite{bib-VAN1996}, decrease effects
of the Oersted field in magneto-resistive or spin-torque oscillator pillars, or more recently boost the current-induced domain-wall propagation speed in nanostripes\cite{bib-PIZ2009,bib-CHA2011}.

In practice \SyFs are used as elements of finite lateral size. It has been shown\cite{bib-SUN2001} and
it is widely used \cite{bib-WOR2004,bib-FUJ2005} that for flat and magnetically soft nanomagnets of
lateral size smaller than a few hundreds of nanometers, the macrospin approximation~(uniform
magnetization) is largely correct. In this framework the coercive field equals the anisotropy field
$2K/\muZero\Ms$ and the energy barrier $K V$ ($V$ is the volume of the dot) preventing spontaneous
magnetization reversal equals the magnitude of anisotropy of the total magnetic energy $\mathcal{E}$,
to which all the physics therefore boils down. Elongated dots are often used to induce or contribute to
an easy axis of magnetization and an energy barrier $\Delta\mathcal{E}$, based on dipolar energy.
Dipolar energy is a quadratic form and thus it is fully determined by its value along the two main in-plane
axes. For single elements $\Delta \mathcal{E}_\mathrm{d}=\Kd V \Delta N$ with $\Delta N$ the difference
between the two in-plane demagnetizing coefficients, and $\Kd=(1/2)\muZero\Ms^2$.

Analytical formulas have been known for a long time to evaluate the mutual energy of an arbitrary set of prisms\cite{bib-RHO1954}. However while simple expressions for $N$ and thus $\Delta\mathcal{E}_\mathrm{d}$ have been described, displayed and discussed for single-layer flat elements\cite{bib-RHO1954,bib-AHA1998}, the analytical expressions and the evaluation of $\mathcal{E}_\mathrm{d}$ in \SyFs have not been discussed in detail so far. Instead the studies requiring estimation of the dipolar energy in SyF, mainly pertaining to MRAM cells\cite{bib-WOR2004,bib-FUJ2005,bib-HAN2008}, have in the best case made use of an
effective so-called attenuation coefficient with respect to self-energy\cite{bib-FUJ2005}, which
requires a numerical evaluation\cite{bib-HAN2008}. The meaning and scaling laws of this attenuation coefficient have never been discussed in detail, hiding the physics at play.
As thermal stability, coercivity and toggle switching fields\cite{bib-WOR2004,bib-FUJ2005,bib-KIM2007c} depend crucially on the interlayer magnetostatic coupling, it is desirable to have a simple yet accurate analytical expression for interlayer dipolar fields. In this manuscript we report exact analytical expressions for the magnetostatics
of \SyFs uniformly-magnetized in each sub-layer. From the numerical evaluation we discuss the physics at play, while from the analytical formulas we propose an approximate yet accurate scaling law for their straightforward table-top evaluation.

\begin{figure}
  \begin{center}
  \includegraphics[width=80mm]{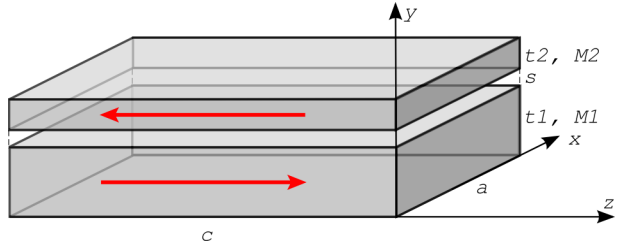}%
  \caption{\label{fig-geometry}Geometry and notations of a prismatic SyF element comprising two
ferromagnetic layers $\mathrm{F}_1$ and $\mathrm{F}_2$.}
  \end{center}
\end{figure}

We first consider \SyF prisms and name F1 and F2 the two ferromagnetic
layers\bracketfigref{fig-geometry}, with magnetization aligned along $z$. This covers the case of both finite-size prisms as well as infinitely-long stripes with a rectangular cross-section. We apply formulas expressing the interaction between two parallel charged surfaces\cite{bib-RHO1954}, and adopt the convenient
notation of $F_{ijk}$ functions, the $i$, $j$ and $k$-fold indefinite integrals along $x$, $y$ and $z$
of the Green's function $F_{000}=1/r$\cite{bib-HUB1998b}. The only such function needed here is

\begin{equation}
  F_{220}=\frac12[x(v-w)L_x+y(u-w)L_y]-xyP_z+\frac16r(3w-r^2)
\end{equation}

\noindent with $u=x^2$, $v=y^2$, $w=z^2$, $r=\sqrt{u+v+w}$, $L_x=(1/2)\ln[(r+x)/(r-x)]$ \etc,
$P_x=x\arctan(yz/xr)$ \etc, and $L_x=0$ and $P_x=0$ for $x=0$ \etc.

The integrated magnetostatic energy of a single prismatic element of thickness $t$ is:

\begin{equation}
  \label{eqn-prism}
  \mathcal{E}_\mathrm{d}=\frac{2\Kd}{\pi}\sum_{\delta_a,\delta_t,\delta_c\in\{0,1\}}(-1)^{\delta_a+\delta_t+\delta_c}
  F_{220}(a\delta_a,t\delta_t,c\delta_c)
\end{equation}

\noindent which normalized to $\Kd$ yields the demagnetizing coefficient $N_z$. It can be verified that
\eqnref{eqn-prism} coincides with the explicit formula already known\cite{bib-AHA1998}. The magnetostatic energy of a prismatic \SyF element may be calculated using the same formalism , may be written as:

\begin{eqnarray}
  \nonumber \mathcal{E}_\mathrm{d} &=& K_{\mathrm{d,1}} N_z(a,t_1,c)V_1 + K_{\mathrm{d,2}}
N_z(a,t_2,c)V_2\\ & & +
2\sqrt{K_{\mathrm{d,1}}K_{\mathrm{d,2}}}N_\mathrm{m}(a,t_1,s,t_2,c) \sqrt{V_1 V_2}
\end{eqnarray}

\noindent with $N_\mathrm{m}(a,t_1,s,t_2,c)= \frac{1}{\pi a \sqrt{t_1 t_2}c}\sum_{\delta_1,\delta_2,\delta_a,\delta_c\in\{0,1\}}(-1)^{\delta_1+\delta_2+\delta_a+\delta_c}
    \times  F_{220}(a\delta_a,s+t_1 \delta_1 +t_2\delta_2,c\delta_c)$ is a mutual magnetostatic coefficient with a negative value, and $V_i=at_i c$ (resp. $K_{\mathrm{d,i}}$) is the volume (resp. dipolar constant) of each single prism~$i$. This equation of dipolar energy is a quadratic form of $M_1$ and $M_2$, generalizing the definition of demagnetizing coefficients.

\begin{figure}
  \begin{center}
  \includegraphics[width=87mm]{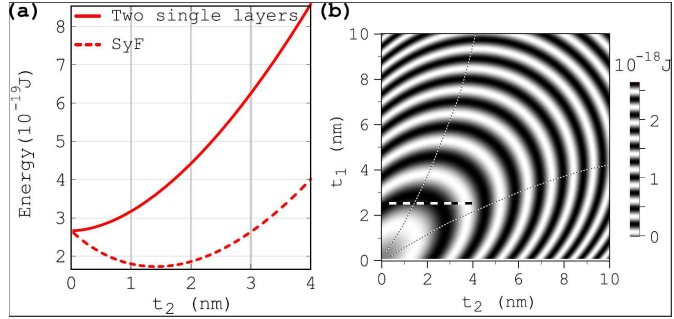}%
  \caption{\label{fig-buildSyF}Magnetostatic energy of a SyF with $c=2a=\lengthnm{100}$,
$M_{1,2}=\unit[10^6]{A/m}$, $s=\lengthnm{0.7}$. (a)~Sum of the energies of two prisms without mutual
interaction, and when embedded in the SyF geometry. $t_1$ is kept constant at $\lengthnm{2.5}$, while
$t_2$ is varied. (b)~Energy of the general SyF. The curved lines are those of minimum energy for either
constant $t_1$ or $t_2$. The thick horizontal dotted line highlights the path for the SyF curve shown in~(a).}
  \end{center}
\end{figure}

\figref{fig-buildSyF}(a) shows $\mathcal{E}_\mathrm{d}$ upon building a \SyF via the progressive
thickness increase of F2 above F1, considering or not the interaction between the two layers. In the
latter case the energy increase nearly scales with $t_2^2$, which is understandable because it is a
self-energy (in $\mathrm{F}_2$ alone). In the coupled case (a SyF) $\mathcal{E}_\mathrm{d}$ retains like for the uncoupled case an overall close-to-parabolic convex shape as can be verified with fitting, however with an initial negative slope. This can be understood as for low $t_2$ the extra edge charges
induced by an infinitesimal increase $\delta t_2$ mainly feel the stabilizing stray field arising from
F1, while for large $t_2$ they feel more the nearby charges induced by F2 itself. Notice that, contrary
to what could be a first guess, the minimum of $\mathcal{E}_\mathrm{d}(t_2)$ occurs before the
compensation of moment~($t_1=t_2$). This stems from the same argument as above, which is that magnetic
charges at an edge of F2 are closer to another than to the charges on the nearby edge of F1, thus
for an identical amount $\delta t_2$ contribute more to $\mathcal{E}_\mathrm{d}$.

\figref{fig-buildSyF}(b) shows the full plot of $\mathcal{E}_\mathrm{d}(t_1,t_2)$ for
$s=\lengthnm{0.7}$. The above arguments appear general. From this figure let us outline three take-away
messages. 1.~For a given $t_1$ the minimum of $\mathcal{E}_\mathrm{d}$ of a SyF is found for
$t_2\gtrsim t_1/2$.  2. At this minimum $\mathcal{E}_\mathrm{d}$ is reduced by only
$\approx\unit[20\mathrm{-}30]{\%}$ with respect to a single-layer element of thickness $t_1$ considered alone.
3.~$\mathcal{E}_\mathrm{d}$ roughly regains the value of the single layer at the moment compensation point ($t_1=t_2$).

This sheds light on results previously noticed empirically, however whose origin and generality had not been highlighted. Wiese \etal reported that the effective dipolar field
anisotropy of a SyF \textsl{basically scales with the inverse net moment}\cite{bib-WIE2005}, \ie like
the inverse strength of Zeeman energy. This suggests that $\Delta\mathcal{E}_\mathrm{d}$ is essentially
independent of the imbalance of moment, which goes against the widespread belief that magnetostatic
energy nearly vanishes upon moment compensation. Our results clarify and quantify this: the dipolar
energy does not differ more than \unit[20-30]{\%} from that of a single layer for $t_2\lesssim
t_1$~(\subfigref{fig-buildSyF}a, dotted line). Saito \etal also reported that the thermal stability of
$\mathrm{Co}_{90}\mathrm{Fe}_{10}[3]~/ \mathrm{Ru}[0.95]~/\mathrm{Co}_{90}\mathrm{Fe}_{10}[5]$ is
similar to that of $\mathrm{Co}_{90}\mathrm{Fe}_{10}[3]$. As explained in the introduction, we recall that thermal stability is determined by the energy barrier along the hard axis direction, with respect to the easy axis direction. In the case of anisotropy arising from dipolar energy and an elongated shape of the element, this barrier can be evaluated straightforwardly by calculating once $\mathcal{E}_\mathrm{d}$ along the \textsl{short} edge of the dot, and second along the \textsl{long} edge of the dot. Doing this we explain the findings of Saito \etal.,
whereas a reduction of $\unit[50]{\%}$ would be expected on the basis of compensated moments~(the
numbers in brackets are thicknesses in nanometers). Our calculations may also be applicable to the
cross-over of vortex versus single domain in flat disks\cite{bib-COW2000} or vortex versus transverse
domain walls in stripes\cite{bib-MIC1997}, whose scaling law $t\times a = \mathrm{Cte}$ may be derived
qualitatively by equaling the energy of a vortex $\sim t$ and that of a single-domain $\sim a^2 t
(t/a)$ (here $V=a^2 t$ is the volume, and $t/a$ the demagnetizing coefficient). Interestingly Tezuka
\etal noticed that \textsl{there is an optimum ferromagnetic film thickness at which SyAF can obtain a
single-domain structure}. This minimum (related to a minimum of demagnetization energy) is found for an
\textsl{imbalanced} thickness in good quantitative agreement with \subfigref{fig-buildSyF}b.

\begin{figure}
  \begin{center}
  \includegraphics[width=87mm]{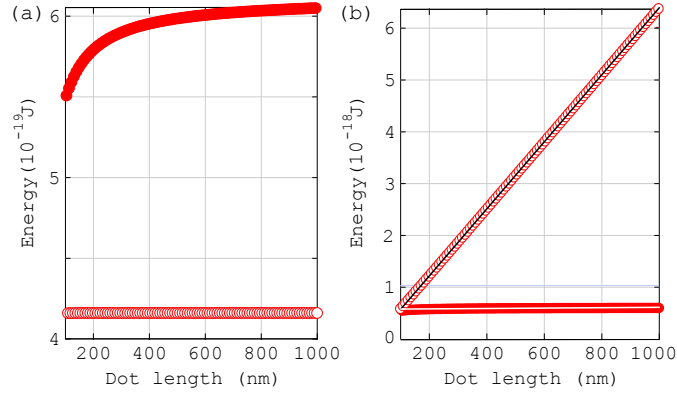}%
  \caption{\label{fig-edgeEnergy}(a)~Energy of a single layer~(full symbols) and SyF~(open symbols)
as a function of dot length, \ie along $z$, while $a=\lengthnm{100}$. (b)~Energy of a single layer as a
function of width (along~$x$, open symbols), and length (along~$z$, full symbols, same curve as in a),
while the other in-plane dimension is kept constant at $\lengthnm{100}$. The lines are linear fits. For
both plots the parameters are: $M_i=\unit[10^6]{A/m}$, $t_1=t_2=\lengthnm{2.5}$, $s=\lengthnm{0.7}$. }
  \end{center}
\end{figure}

With a view to promote the use of accurate magnetostatics for SyF while eliminating the need for
numerical evaluation, we derived approximate yet highly accurate expressions for
$\mathcal{E}_\mathrm{d}$. \subfigref{fig-edgeEnergy}a shows that to a very good approximation,
$\mathcal{E}_\mathrm{d}$ is proportional to the width of the element (along~$x$) and is independent of
its length (along~$z$). This is already accurate for a single-layer~($t_2=0$), and is very accurate
close to the compensation $t_1=t_2$ because edges then behave as lines of dipoles, whose stray field quickly
decays with distance~($\sim 1/r^2$). Thus $\mathcal{E}_\mathrm{d}$ boils down to a single line integral
along its edge:

\begin{eqnarray}
  \label{eqn-edgeIntegral}
  \mathcal{E}_\mathrm{d} & = & E_\lambda \oint (\vect m .\vect n)^2 \diff{s} \\
              & = & E_\lambda  \oint \frac{|\diff x|}{\sqrt{1+(\partial_x f)^2}} \\
              \label{eqn-edgeIntegral3}
              & = & E_\lambda  \int_0^{2\pi} \frac{(r\sin\theta-\partial_\theta
r\cos\theta)^2}{\sqrt{(\partial_\theta r)^2+r^2}}\diff\theta
\end{eqnarray}

\begin{figure}
  \begin{center}
  \includegraphics[width=85.847mm]{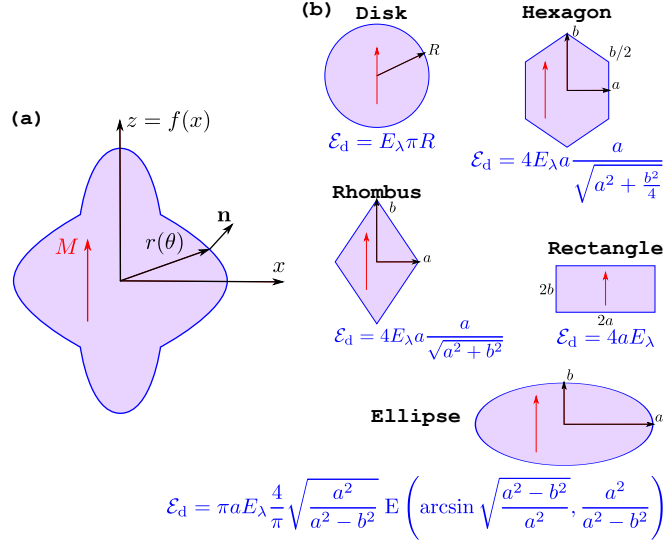}%
  \caption{\label{fig-shapes}(a)~notations for the calculation of edge energy (b)~integrated
magnetostatic energy for various shapes. $\mathrm{E}$ is the elliptical integral of the second kind.
See text for the definition of $E_\lambda$.}
  \end{center}
\end{figure}

\eqnref{eqn-edgeIntegral} is the general expression, expressed in the following two lines in cartesian
and polar coordinates\bracketsubfigref{fig-shapes}a. $E_\lambda$ is the density of magnetostatic energy
per unit length of edge, a concept once discussed in the case of single layers\cite{bib-YUA1992}.
Equations (\ref{eqn-edgeIntegral}-\ref{eqn-edgeIntegral3}) apply to an arbitrary shape of perimeter~(not simply
rectangles for prisms) by considering the in-plane angle $\varphi$ between magnetization and the normal to the edge.
It can be verified that for a \SAF we have, with an accuracy better than $\unit[10]{\%}$ for geometrical
parameters relevant for practical cases, \ie $t_{1,2}$ in the range of $\lengthnm{2-10}$ and
$s in the range of \lengthnm{0.5-1}$:

\begin{equation}
  \label{eqn-simpleEdgeEnergy}
  E_\lambda\approx(1/2)\Kd t^2
\end{equation}

The meaning of \eqnref{eqn-simpleEdgeEnergy} is straightforward: due to the short range of interaction
between dipolar lines, the density of dipolar energy is non-zero only in the vicinity of the edges,
with a lateral range $t$. Thus a volume $t^2$ is concerned with a line density of energy of the order
of $\Kd$. Expressions for non-compensated cases~(including single elements) may also be evaluated. This
provides us with analytical expressions for the magnetostatics of \SyFs for the most usual
shapes\bracketsubfigref{fig-shapes}b.

A scaling law sometimes used as a first guess is based on the point dipole approximation. In this framework the energy gained by coupling $\mathrm{F}1$ and $\mathrm{F}2$ would roughly scale with $\Kd a^4/t$, resulting from two point moments $\Ms a^2t$ interacting like $1/t^3$~(for $s\ll t$, and assuming lateral dimensions of the order of $a$).
The scaling arising from our exact calculation is $a E_\lambda\sim\Kd at^2$~(\eqnref{eqn-simpleEdgeEnergy} and \figref{fig-shapes}a). The point dipole approximation is thus clearly incorrect with an extra scaling $(a/t)^3$~(see
\subfigref{fig-shapes}a) which largely overestimates the dipolar coupling. This is a general argument for any flat element, where dipolar fields are short-ranged\cite{bib-FRU1999b} and thus the point-dipole approximation is clearly incorrect.

To conclude we derived exact formulas for the magnetostatics of prism \SyF, and simple yet accurate
forms for \SyFs of arbitrary shapes. These simple forms may be used straightforwardly to derive scaling
laws for all aspects of SyF physics pertaining with dipolar energy such as thermal stability,
coercivity and anisotropy field. Notice that similar to the case of single flat elements edge roughness is liable to reduce significantly dipolar energy\cite{bib-COW2000c,bib-UHL2004b,bib-MOO2009}, so that the theoretical predictions need to be considered as upper bounds to the experimental values. The non-uniformity of magnetization is not expected to have a significant impact for lateral sizes below a few hundreds of nanometers\cite{bib-SUN2001}.

We acknowledge useful discussions with Y. Henry, IPCMS-Strasbourg.


\end{document}